\documentclass{article}
\usepackage{graphicx} 
\usepackage{xcolor} 
\usepackage{amssymb,mathtools}
\usepackage{amsthm}
\usepackage[a4paper, total={6in, 8in}]{geometry}
\usepackage{algorithm}
\usepackage{algpseudocode}
\usepackage{subcaption}
\usepackage{amsmath}
\usepackage{lineno}
\usepackage{soul}
\usepackage{hyperref}

\newcommand{\polygonal}[1]{$#1$-\textit{path}}
\newcommand{\polygonals}[1]{$#1$-\textit{paths}}
\newcommand{\pair}[2]{$\langle #1, #2 \rangle$}

\newcommand{\abs}[1]{\lvert#1\rvert}

\newtheorem{definition}{Definition}
\newtheorem{theorem}{Theorem}
\newtheorem{lemma}{Lemma}

\graphicspath{ {images/} }

\title{Computing Optimal Trajectories for a Tethered Pursuer in Straight-Line Motion \footnote{Supported in part by grants PID2020-114154RB-I00 and TED2021-129182B-I00 funded by MCIN/AEI/10.13039/501100011033 and the European Union Next\allowbreak{GenerationEU/PRTR}.}}
\author{A. Barrera-Vicent$^{1}$, J.M. D\'iaz-B\'añez$^{1}$, F. Rodr\'iguez$^{1*}$ and V. Sánchez-Canales$^{1}$
\thanks{$^{1}$ A. Barrera-Vicent ({\tt\small aurebv@gmail.com}), J.M. D\'iaz-B\'añez, F. Rodr\'iguez and V. Sánchez-Canales ({\tt\small [dbanez,frodriguex,vscanales]@us.es}) are with the Department of Applied Mathematics II, University of Seville, Spain.}
\thanks{$*$ Corresponding Author}
}

\date{}

\begin{document}

\maketitle

\begin{abstract}
In this paper, we introduce a trajectory planning problem for a marsupial robotics system consisting of a ground robot, a drone or aerial robot, and a bounded-length taut tether $L$ connecting the two robots. 
We study a scenario in which the ground robot and the drone move along parallel lines within a vertical plane. The drone follows a predefined back and forth trajectory at a constant speed, and the problem is to determine an optimal path for a ground robot 
The objective is to calculate a minimum link trajectory, a back and forth path composed of the fewest possible direction changes, and a constant speed for the ground robot, ensuring that the separation between the two robots does not exceed $L$ at any point.
This problem can be framed within the context of a pursuit-evasion game, where the evader's trajectory is known, and the goal is to compute an optimal trayectory for the pursuer. Employing geometric modeling techniques, we develop an optimal algorithm to compute a parameterized minimum-link trajectory for the ground-based pursuer, given the a priori known trajectory of the aerial evader. Additionally, we solve three interconnected geometric optimization problems by systematically exploiting their inherent relationships.
\end{abstract}


\section{Introduction}

In applications such as urban search and rescue, a novel multi-robot system known as the ``marsupial multi-robot'' has been developed \cite{murphy1999marsupial}. This system consists of a ground robot that acts as a mobile base station for an aerial robot. The term marsupial robot draws inspiration from nature, evoking the image of a larger ``mother'' robot carrying one or more smaller ``offspring'' robots, similar to how a kangaroo mother carries her young in a pouch.

In this paper, we study an optimization problem inspired by path planning tasks with a tethered marsupial robotics system (composed of a ground robot, an aerial robot or drone, and a extensible taut tether of maximum length $L$ connecting the robots).  A marsupial multi-robot system is an alternative 
to extend the flight duration of the drone, in which one robot carries and can deploy one or several other robots \cite{murphy2000marsupial,hourani2011marsupial}. Path planning in tethered robotics systems has gained interest addressing challenges in various applications, particularly in
confined or hazardous environments.
See \cite{martinez2023path,fabio} for an example of the path-planning problem of such multi-robot systems. 

In a path-planning problem involving a marsupial robotic system, two synchronized trajectories must be computed: one for an unmanned aerial vehicle (UAV or drone) and another for an unmanned ground vehicle (UGV). Given the UAV's predefined path, the UGV serves as a mobile base, ensuring that the distance between the two robots does not exceed the tether's length, $L$.

In our study, we consider a particular scenario. We are provided with a predefined trajectory and a constant speed for the UAV. The objective is to design an algorithm that computes an optimal trajectory and a constant speed for the ground vehicle, ensuring that the distance between the two robots does not exceed a specified limit $L$. The cost of the ground robot's trajectory is determined by the number of turns it makes. Additionally, we solve optimization problems where the cost is defined by either the speed of the ground robot or the total length of its trajectory.

In this scenario, we study the following 2D-version: 
both the drone and the ground vehicle move within a vertical plane along parallel lines $l$ and $l'$, respectively, while maintaining a maximum distance of $L$ between them via an extensible tether. The drone moves forward or back along the line $l$ using a given constant speed. The problem is to calculate a constant speed and a walk along the line $l'$ so that the ground vehicle follows the drone at a distance of at most $L$, minimizing the number of
turns (links or changes of direction). We solve the problems using geometric optimization tools. By leveraging properties of an optimal solution and utilizing geometric structures, such as the convex hull of a set of points, we propose a linear-time algorithm to solve the problem.

The problem is particularly relevant in applications where maintaining proximity between two agents is crucial, such as in environments with communication constraints or power supply limitations. 
In the case of tethered systems, by ensuring that the ground vehicle's movements are both efficient and synchronized with the drone, our approach enhances the operational effectiveness of marsupial robotic systems.
The problem is also of interest in computer vision, where visibility-based target tracking has received
a lot of attention in the research community. In this context, the goal of the observer is to maintain a persistent line-of-
sight with the target.


The remainder of the paper is organized as follows. In Section \ref{sec:RW}, we discuss related
works. Section \ref{sec:GF} presents a geometrical formulation of the main problem. A solution for the minimum-speed problem is described in Section \ref{sec:slope}. Section \ref{sec:min-link-path} shows a linear-time algorithm for computing optimal minimum-link and minimum-length ground vehicle trajectories.
The paper concludes with a discussion of key findings and proposed future research directions in Section \ref{sec:conclusion}.

\section{Related work}\label{sec:RW}

The problem investigated in this work is motivated by trajectory planning in marsupial robotic systems, where the robots' motion is constrained at all times by the finite length of the tether connecting them. However, the problem can be posed in the context of pursuit-evasion games, where the \emph{observer} is the ground robot and the \emph{target} is the drone. 
Pursuit-evasion games are highly relevant in robotics, as many applications can be modeled using this framework.
Most existing approaches to visibility-based pursuit-evasion problems focus on pursuer strategies that ensure continuous visual contact with the evader, under the assumption that both share the same upper speed limit \cite{lavalle1997motion,robin2016multi,frew2003trajectory}.
Our problem can be labeled as a target tracking problem. Target tracking involves the challenge of planning the motion for a mobile observer to effectively follow a moving target while navigating around obstacles. The goal is to ensure that the observer maintains visibility or proximity to the target. This problem requires dynamic path planning, real-time decision-making, and often incorporates constraints such as maintaining a safe distance, avoiding collisions, and optimizing the observer's trajectory for efficiency. Most of the problems assume that observer knows the current position of the
target as long as the target is in the observer’s line-of-sight \cite{zou2018optimal}.
In these scenarios, the observer is equipped with a vision sensor to track the target, while the environment may contain obstacles that occlude the observer's view of the target
\cite{bhattacharya2016visibility}.

In a pursuer-evasion setting, several solution concepts can
be used to compute the optimal strategy of the pursuer. 
For example, minimal length paths and time-optimal trajectories
have been obtained for robots with different dynamic and
kinematic configurations \cite{balkcom2002time}, minimum
wheel-rotation paths are presented in \cite{chitsaz2009minimum}, paths minimizing the
energy consumption are studied in \cite{tokekar2014energy}. An optimal strategy for the pursuer to maintain
a constant distance with the evader at minimal velocity is presented in \cite{murrieta2011tracking}.

On the other hand, it is well known that many problems in computational geometry have been inspired by path planning in robotics. Typical objective functions are minimizing or maximazing the length of the trajectory or minimizing the number of links. 
The minimum-link path problem is fundamental in computational geometry \cite{suri1986linear,mitchell1992minimum,mitchell2014minimum}. The term links is used  for edges of the path and bends for the vertices. The problem addresses the following question: Given a polygonal domain $D$ and two points $s$ and $t$ within $D$, what is the polygonal path connecting $s$ to $t$ that lies entirely within $D$ and consists of the fewest possible links? 
Due to their diverse applications, many different variants of minimum-link paths have been considered in
the literature.
See \cite{maheshwari2000link} for an excellent review of min-link
paths.

There exists a classical distinction for both shortest and min-links paths: simple polygons vs. polygonal domains. The
former are a special case of the latter: simple polygons are domains without holes. Many problems admit
more efficient solutions in simple polygons. For example,   the
shortest path problem can be solved in $O(n)$ and $O(n\log n)$ time for simple polygons and polygonal domains, respectively \cite{guibas1986linear, hershberger1994computing}. For minimum-link paths, $O(n)$-time algorithms are known for simple polygons
\cite{suri1986linear}, but for polygonal domains with holes the fastest known algorithm runs in nearly quadratic time \cite{mitchell1992minimum}. In this work, we show how the pursuit problem for a ground robotic agent can be formulated as a minimum-link path problem within polygonal environments under motion constraints.

\section{Geometric Formulation}\label{sec:GF}

\begin{figure}[H]
\begin{subfigure}{.43\linewidth}
  \includegraphics[width=\linewidth]{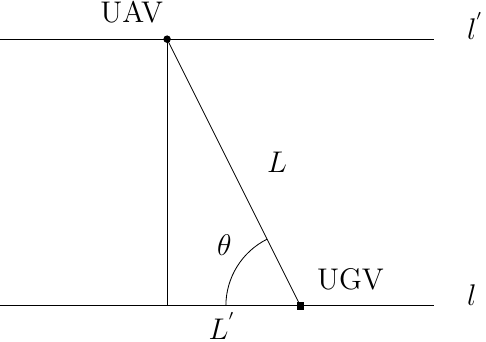}
  \caption{}
  \label{fig:parallel-lines}
\end{subfigure}\hfill 
\begin{subfigure}{.5\linewidth}
  \includegraphics[width=\linewidth]{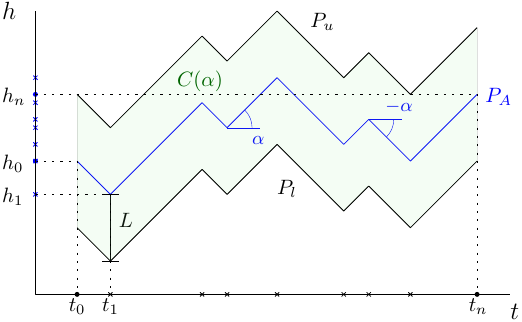}
  \caption{}
  \label{fig:alpha-path}
\end{subfigure}
\caption{(a) UAV and UGV move in parallel lines. (b) Geometric formulation.}
\end{figure}

In this section, we formulate the problems under consideration in a geometric setting. We consider two vehicles: an aerial robot (UAV) and a ground robot (UGV), both modeled as points moving at constant speeds (back and forward) along parallel straight lines $l$ and $l'$, respectively, within a vertical plane. The ground robot, $G$, must remain, at any time, within a maximum distance $L$ from the aerial robot, $A$.  
Since $l$ and $l'$ are parallel, the vertical distance between $G$ and $A$ remains constant. Therefore, for simplicity, we can identify $A$ with its orthogonal projection onto line $l'$ by changing the maximum allowed distance $L$ by $L'=L\cos{\theta}$ as in Figure \ref{fig:parallel-lines}. A path for $A$ on $l'$ is defined by the set of \emph{turn points} $P_A = \{h_0, h_1, \dots, h_n\}$ ordered by time. 
Thus, given a path for $A$, denoted by $P_A$, where $A$ moves at a constant speed $\alpha$, the problem is to compute a path on $l'$ for the ground vehicle $G$, denoted by $P_G$, such that $G$ moves at a constant speed $\beta$, maintain a distance of at most $L$ from $A$\footnote{For ease of notation we rename $L'=L$, even though $L'<L$
.} during a period of time $[t_0,t_n]$, 
and minimizes the number of turns (changes of direction). 

The problem admits a simple geometric formulation by representing the path of $A$ in a space-time plane, as illustrated in Figure \ref{fig:alpha-path}. The path $P_A$ can be represented in the space-time plane by a polygonal line with vertices
$\{(t_0, h_0), (t_1, h_1), \dots, (t_n, h_n)\}$. The first components $t_i$ represent the times at which the vehicle reaches the values $h_i$ in $P_A$. The links of $P_A$ alternate between the slopes $\alpha$ and $-\alpha$. 
In this work, with slight abuse of notation, we also denote by $\alpha$ the angle between a path segment and the positive $x$-axis.

A path with this property is called an $\alpha$\emph{-path} and an
$\alpha$\emph{-corridor}, $C(\alpha)$, is defined by the locus of points whose vertical distance is at most $L$ from an $\alpha$-path $P_A$. We say that a $\beta$-path is \emph{feasible} if it traverses the interior of the corridor $C(\alpha)$. 

Based on these definitions, we can reformulate our optimization problem in a geometrical setting as follows:

\vspace{.25cm}
\emph{Min-Link Problem: Given an $\alpha$-corridor, compute a minimum-link feasible $\beta$-path}.
\vspace{.25cm}

An immediate intuition is that a feasible $\beta$-path with minimum $|\beta|$ provides the solution to the problem. In fact, a solution with $\beta=0$ corresponds to a horizontal line crossing the $\alpha$-corridor, which is the feasible $\beta$-path with the minimum number of links and, moreover, the minimum length. See an example in Figure \ref{fig:alpha-path}.

In the following, we focus on the problem of computing the minimum absolute slope $\beta$ for a feasible $\beta$-path. Afterward, we show how to use this minimum-slope $\beta$-path to construct a minimum-link $\beta$-path.
This means that, in the application from which this problem arises, minimizing the speed of the ground vehicle also results in minimizing its number of turns. The optimization problem is the following.

\vspace{.25cm}
\emph{Min-Slope Problem: Given
an $\alpha$-corridor,
 calculate the minimum value of $|\beta|$  for a feasible $\beta$-path}.
\vspace{.25cm}

In this paper, we also solve a third problem.

\vspace{.25cm}
\emph{Min-Length Problem: Given an $\alpha$-corridor, compute a minimum-length feasible $\beta$-path}.
\vspace{.25cm}

\section{Finding the minimum feasible slope}\label{sec:slope}

\begin{figure}
    \centering
    \includegraphics[scale=1.1]{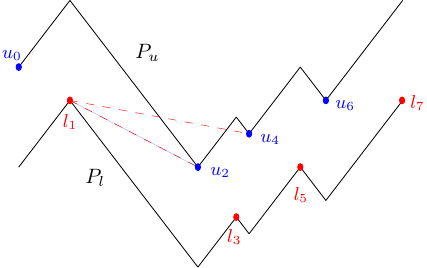}
    \caption{Lower and upper reflex points. The pair \pair{l_5}{u_4} is not an overlapping pair, whereas the pair \pair{l_1}{u_2} is. The mentioned pairs are visible, but the pair \pair{l_1}{u_4} isn't.}
    \label{fig:overlapping_example}
\end{figure}

An $\alpha$-corridor is bounded by two \polygonals{\alpha}, denoted by $P_l =\{l_0, \dots, l_n\}$ and $P_u=\{u_0, \dots, u_n\}$, which are the vertical translations of the \polygonal{\alpha} $P_A$ downward and upward, respectively, by a distance $L$. See Figure \ref{fig:alpha-path}. We refer to $C[i,j]$ to the section of $C(\alpha)$ between $t_i$ and $t_j$ for $0\leq i < j \leq n$.
A \emph{reflex} point is a vertex in the boundary of the corridor where the internal angle exceeds 180 degrees. Figure \ref{fig:overlapping_example} illustrates the following definition.

\begin{definition}
    Let \pair{l_i}{u_j} be a pair  of reflex points, where $l_i\in P_l$ and $u_j\in P_u$: 
\begin{itemize}
       \item [1)] We say that \pair{l_i}{u_j}  \emph{overlaps} if $u_j$ lies below $l_i$. 
    \item [2)] \pair{l_i}{u_j} is a  \emph{visible
    pair}  if the segment connecting $l_i$ and $u_j$, denoted by $\overline{l_iu_j}$, lies entirely within the interior of the corridor. 
    \item [3)] The \emph{slope} of the pair \pair{l_i}{u_j}, denoted by $s(l_i,u_j)$, is the absolute value of the slope of the segment $\overline{l_iu_j}$.
\end{itemize}   
\end{definition}


Notice that if there are no pairs of reflex points that overlap in the corridor, then the solution to the Min-Slope Problem is $\beta=0$, which means that a horizontal line is a traversal of the corridor. 

The following Lemmas support the first result of the paper.

\begin{lemma}    \label{lemma:overlapping}
   Let \pair{l_i}{u_j} be an overlapping pair of reflex points in an $\alpha$-corridor, $l_i\in P_l$ and $u_j\in P_u$.
   \begin{itemize}
       \item [a)] 
        If \pair{l_i}{u_j} is a visible pair, then $s(l_i,u_j)$ provides the minimum slope required to traverse the section $C[i, j]$.
        \item [b)] If \pair{l_i}{u_j} is not a visible pair, then there is an overlapping pair of reflex points, \pair{l_k}{u_m} in $C[i, j]$ with a bigger slope than $s(l_i,u_j)$.
   \end{itemize}

\end{lemma}

\begin{proof} 
    Figure \ref{fig:overlapping_example} illustrates both parts of Lemma \ref{lemma:overlapping}: (a) as \pair{l_1}{u_2} is a visible and overlapping pair, the minimum slope needed to cover the corridor between both is $s(l_1,u_2)$; (b) the overlapping pair \pair{l_1}{u_4} is not visible, then there exist other overlapping pairs, for instance,  \pair{l_1}{u_2}, with $s(l_1,u_2)>s(l_1,u_4)$.
\end{proof}



\begin{lemma}\label{lemma:feasible-slopes}
Let $C(\alpha)$ be an $\alpha$-corridor, and let $\beta$ and $\delta$ be two slopes such that $|\delta| \leq |\beta| \leq |\alpha|$. If there exists a feasible $\delta$-path, then there exists a feasible $\beta$-path.
\end{lemma}

\begin{proof}
Consider a segment $\overline{DE}$ with slope $\delta$ within $C(\alpha)$. Let $p_1, \ldots, p_k$ be the projection points of reflex vertices along this segment. As illustrated in Figure \ref{fig:slopes-proof}, it is easy to construct a \polygonal{\beta} path that travels from $D$ to $p_1$, then from $p_i$ to $p_{i+1}$ for each $i<k$, and finally from $p_k$ to $E$, all while remaining entirely within the corridor. By applying this construction to every segment of a feasible \polygonal{\delta}, we can build a corresponding feasible \polygonal{\beta}.
\begin{figure}[ht]
    \centering
\includegraphics[scale=0.7]{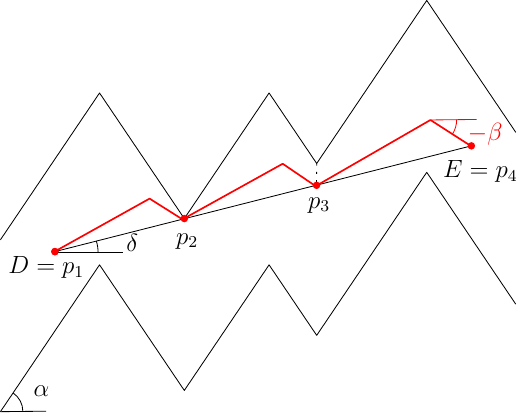}
    \caption{Given a segment $\overline{DE}$ with slope $\delta$ within $C(\alpha)$, we show how to build a feasible \polygonal{\beta} from $D$ to $E$ when $\abs{\delta} \leq \abs{\beta} \leq \abs{\alpha}$. }
    \label{fig:slopes-proof}
\end{figure}
\end{proof}

\begin{theorem}\label{theo:min-slope}
The minimum slope $\beta$ required to traverse $C(\alpha)$ using a feasible $\beta$-path is determined by the largest slope among all overlapping pairs of reflex points.
\end{theorem}

\begin{proof} 
    Let \pair{l_i}{u_j} be the overlapping pair of reflex points with the maximum slope $\beta^*=s(l_i,u_j)$. 
    From Lemma \ref{lemma:overlapping} b), \pair{l_i}{u_j} is a visible pair. From Lemma \ref{lemma:overlapping} a), no slope lower than $\beta^*$ is feasible because would not be feasible in the section $C[i, j]$.
    As the pair \pair{l_i}{u_j} defines the maximum slope among all pairs of overlapping reflex points, any other section in the corridor would require a minimum absolute slope lower than or equal to $\beta^*$. According to Lemma \ref{lemma:feasible-slopes}, the slope $\beta^*$ will be feasible for any section of the corridor and the result follows. 
\end{proof}

Theorem \ref{theo:min-slope} enables a straightforward $O(n^2)$-time algorithm to compute the minimum slope we are looking for. This algorithm iterates through each pair of reflex points \pair{l_i}{u_j}, calculates their slope (when they overlap), and retains the largest slope encountered. 
If no pairs overlap, a horizontal line provides the optimal slope. In what follows, we present a linear-time solution to the Min-Slope Problem using a more sophisticated geometric approach.

\subsection{An optimal algorithm}

\begin{figure}[htb]
    \centering
\includegraphics[scale=.8]{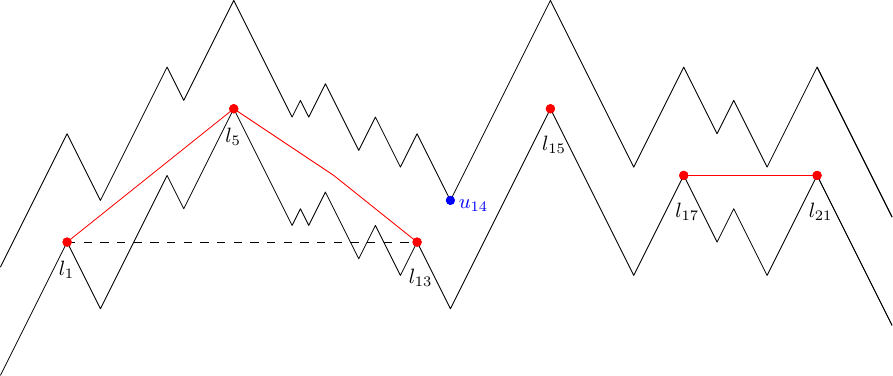}
    \caption{Illustration of three Maximum Convex Chains (MCCs) along a corridor computed from left to right, shown in red: $MCC[1]= \{l_1, l_5, l_{13}\}$, $MCC[15]=\{l_{15}\}$, and $MCC[17]=\{l_{17}, l_{21}\}$.}
    \label{fig:mccs}
\end{figure}

\begin{definition} 
Let $P_l[i,j]$ denote the set of reflex points in $P_l$ within section $C[i,j]$. The \emph{Maximum Convex Chain from} $l_i$, denoted by $MCC[i]$, is the subset of vertices in the convex hull of $P_l[i,j]$ that lie on or above the segment $\overline{l_i l_{j}}$, where $l_{j}\in P_l$ is the rightmost reflex point from $l_i$ such that the $\alpha$-path  $P_u$ does not intersect with the polygonal line defined by $MCC[i]$. See Figure \ref{fig:mccs} for an example.
\end{definition}

Our approach leverages the sequence of Maximum Convex Chains (MCCs) of the corridor to address the Min-Slope Problem. Once this sequence is computed, we detail how to determine the optimal slope for each chain within the corridor and, subsequently, compute the overall optimal slope $\beta$ for a $\beta$-path traversing the entire corridor.

\subsubsection{Computing the sequence of MCCs}\label{section-graham}

We use a variant of Graham's scan for polygonal chains similar to \cite{lee1983finding} but ensuring that no reflex vertex in $P_u$ lie below the $MCC$s. See \cite{bronnimann2006space} for a space-efficient approach that can be adapted when an online algorithm is required.

First, we are able to compute $MCC[i]$ in linear time. The method is based on the Graham's algorithm maintaining, in addition, an array $U$ with an upper reflex point for each $l_j\in P_l, j>i$, to ensure the construction of the convex chain is collision free. Specifically, for each $l_j$, the upper reflex vertex $U[j]$ is chosen such that the segment $\overline{l_jU[j]}$ has the lowest slope\footnote{This refers to the slope of a segment and not the absolute slope of a pair.}. This upper reflex point ensures the visibility required to add new points to the convex chain.
The process uses a stack to scan backwards through the reflex points of $P_l$, checking for convexity. If the convexity condition is violated, points are removed from the stack, and $U[j]$ is updated accordingly. When an intersection is detected, the procedure skips to the next reflex point in $P_l$ and resumes.
For example, in Figure \ref{fig:mccs}, the segment $\overline{l_5l_{15}}$ intersects the upper polygonal line, with $U[5]=u_{14}$. Consequently, $l_{15}$ cannot be added to the convex chain $MCC[1]$.
This approach guarantees that the sequence of maximum convex chains (MCCs) is computed in linear time, as the backward scanning and updates are efficient and avoid redundant calculations.
It is important to note that an MCC is unique when computed from left to right. However, computing it from right to left may yield a different MCC. Figure \ref{fig:mccs} illustrates an MCC computed from left to right.

\subsubsection{Solving the Min-Slope Problem}\label{solving min-slope}

Let $l_i$ be the highest point in $MCC[k]$ that ends at $l_j$.  
We propose an efficient algorithm to determine the least steep slope required to traverse the section $C[i, j+1]$ of the corridor. Our method computes the optimal slope for a $\beta$-path originating at $l_i$ with a rightward orientation.  
A similar approach is used to determine the optimal slope that extends backward within an MCC, which requires pre-computing the MCCs from right to left. Finally, by considering the maximum of the forward and backward slopes across all MCCs (computed from both directions) and comparing the resulting values, we obtain the global minimum slope required to solve the Min-Slope Problem.



Let $MCC[i]=\{l_i, \dots, l_{j}\}$, where $l_i$ is the highest point in $MCC[k]$. 
Theorem \ref{theo:min-slope} states that the minimum slope $\beta$ required to traverse the section $C[i, j+1]$ is determined by the steepest slope among all visible overlapping pairs of reflex points within that section. Given any lower reflex point  $l_p\in P_l$ in the interval $C[i,j]$, we define $o_p(j)\in P_u$ as the upper reflex point such that the slope $s(l_p,o_p(j))$ is the maximum among all overlapping pairs \pair{l_p}{u_q}, $u_p\in P_u$, $p\leq q\leq j+1$.

\begin{lemma}\label{lemma:opt-pair}
Let $l_i$ be the highest point in $MCC[k]$. Given $MCC[i]=\{l_i, \dots,l_j\}$, the minimum feasible slope required to traverse the section $C[i,j+1]$ is $s(l_p,o_p(j+1))$, where
$l_p$ is the first point in $MCC[i]$ (from left to right) such that the pair \pair{l_p}{o_p(j+1)} is visible.
\end{lemma}

\begin{proof} 
Let \pair{l_p}{o_p(j+1)} be the first visible overlapping pair of reflex points (from left to right) within $C[i, j+1]$. Using the fact that no upper reflex point lies below $MCC[i]$, 
the point $l_p$ must be in $MCC[i]$. Additionally, Lemma \ref{lemma:overlapping} b) implies that reflex points in $MCC[i]$ to the left of $l_p$ do not determine 
the maximum overlapping pair within $C[i, j+1]$.

Let $r$ be the ray originating from $l_p$ and passes through $o_p(j+1)$. Due to the selection of $o_p(j+1)$, any lower reflex point in $MCC[i]$ to the right of $l_p$ cannot have overlapping reflex points below $r$. 
Thus, the maximum slope within $C[i, j+1]$ is $s(l_p, o_p(j+1))$, and by Theorem \ref{theo:min-slope}, the result follows.
\end{proof}

Lemma \ref{lemma:opt-pair} provides the basis for an efficient procedure to compute the optimal $\beta^*$. By performing a sweep from $l_i$ to $l_j$ on $MCC[i]$, we can solve the problem for the decreasing portion of $MCC[k]$ in $O(j-i)$ time. This leads to the following result:

\begin{theorem} \label{theo:min-slope-complexity}
The Min-Slope Problem  can be solved in $\Theta(n)$ time.
\end{theorem}
\begin{proof}
Calculating the sequence of $MCCs$ of the corridor (both from left to right and from right to left) takes $O(n)$ time. By construction of $MCCs$, the pair of overlapping reflex points with maximum slope must lie within some $MCC[k]$ corresponding to a lower reflex point $l_k$.
After that, computing the minimum feasible slope for all $MCCs$ will take $O(n)$ time in total. We are left with a linear number of candidates for $\beta^*$. As a direct consequence of Lemma \ref{lemma:feasible-slopes}, computing the maximum of these candidates will output the minimum feasible slope to cover the entire corridor. 
\end{proof}

\section{Computing a minimum-link $\beta$-path} \label{sec:min-link-path}
Let $\beta^*=s(l_i,u_j)$ be the minimum feasible slope for a $\beta$-path traversing the corridor. 
In this section, we prove that there exists a feasible \polygonal{\beta^*} solving the Min-Link Problem. 
The following property is a direct consequence of Lemma \ref{lemma:overlapping}:

\begin{lemma}\label{cor:opt-beta}
    Any feasible \polygonal{\beta^*} must include the reflex points $l_i$ and $u_j$.
\end{lemma}

Let $l_i$ and $u_j$, $i < j$, be overlapping reflex points that determine $\beta^*$.  We outline a greedy algorithm to construct a \polygonal{\beta^*}, $B^*$, with the minimum number of links in section $C[i,n]$ of the corridor. A similar approach can be used from $l_i$ backwards to construct the polygonal path for the section preceding $C[i,n]$. From Lemma \ref{cor:opt-beta},  $B^*$ passes through $l_i$ and $u_j$. 

\vspace{.5cm}
{\bf Greedy Construction (GC):} \emph{Starting from $l_i\in B^*$ and using slope $\beta^*$ or $-\beta^*$, advance in the specified direction as much as possible within the corridor. Once further advancement is no longer possible, place a link at the point where the maximum further progress can be made. After placing the link, start again from this point in the opposite slope and so on.}
\vspace{.5cm}

\begin{figure}[htb]
    \centering
    \includegraphics[scale=0.6]{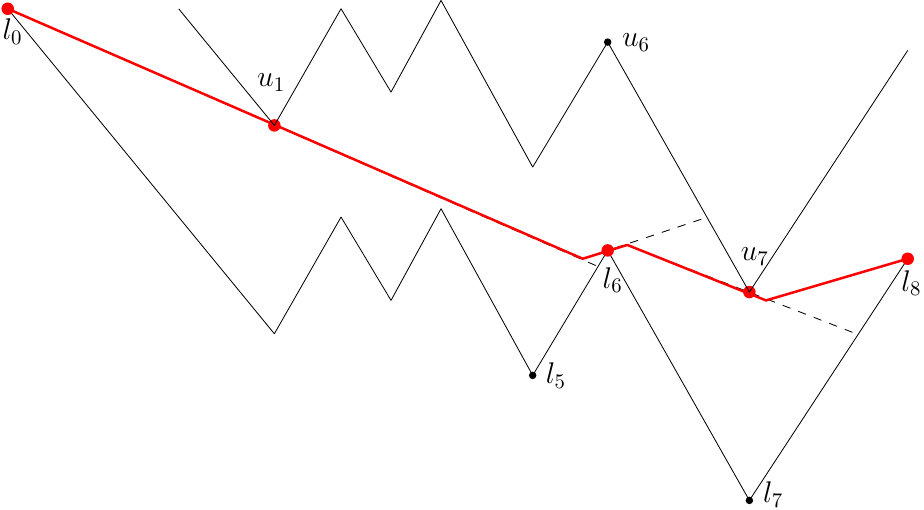}
    \caption{Path constructed using GC from the reflex point $l_0$.}
    \label{fig:path-reflex}
\end{figure}

A \polygonal{\beta^*} constructed using the greedy approach will necessarily include at least one reflex point of the corridor in each of its segments.
Specifically, when the orientation of a segment in $B^*$ is positive, the segment is supported by at least one lower reflex point. Conversely, when the orientation is negative, the segment is supported by at least one upper reflex point. Figure \ref{fig:path-reflex} illustrates the construction. It is straightforward to prove that GC yields a feasible \polygonal{\beta^*} within $C[i,n]$.

The following result guarantees the correctness of our approach.
\begin{theorem}\label{theo:min-links}
    The \polygonal{\beta^*}, constructed by the GC algorithm and denoted as $B^*$, is an optimal path to the Min-Link Problem. 
\end{theorem}

\begin{proof}
If $\beta^*= 0$, we are done. Assume $\beta^*\not= 0$ and let $l_i$ and $u_j$ be the overlapping reflex points providing the optimal slope $\beta^*$. 
We prove the result for section $C[i,n]$.  The same argument can be applied for section 
$C[1,j]$, that is, from the beginning of the corridor until the reflex point $u_j$.


Assume that $B^*$ passes through the reflex points $b^*_1, b^*_2, \dots, b^*_k, b^*_{k+1}, \dots, b^*_m$, $l_i = b^*_1$ and $u_j = b^*_2 $. Define \emph{level} $k$ as the vertical line passing through $b_k$. The proof proceeds by induction on $k$. For a feasible $\beta$-path $B$, where $\beta \geq \beta^*$, let $\text{sign}(B, k)$ denote the sign of the slope of the segment in $B$ oriented from left to right that touches level $k$. Additionally, let $ \vert B_k \vert $ represent the number of links in a polygonal $\beta$-path $B$ up to level $k$.  

We prove by induction the following property that certifies the statement:

\begin{itemize}
    \item[$\bullet$] If $\text{sign}(B, k) =\text{sign}(B^*, k)$, then $\vert B^*_k \vert \leq \vert B_k\vert$.
    \item[$\bullet$] If $\text{sign}(B,k) \neq \text{sign}(B^*, k)$, then $\vert B^*_k \vert \leq \vert B_k\vert +1$. 
\end{itemize}

\underline{Base case ($k=2$):} Set $l_i=b_1, u_j=b_2$.
As \pair{b_1}{b_2} is a visible overlaping pair, $B^*$ reaches level $2$ without any turns. For $B$, if $\text{sign}(B, 2) = \text{sign}(B^*, 2)$, $\vert B^*_2 \vert \leq \vert B_2\vert$. However, if $\text{sign}(B, 2) \neq \text{sign}(B^*, 2)$, $B$ requires at least one additional link in section $C[i,j]$ because \pair{l_i}{u_j} is an  overlapping pair.

\underline{Inductive step ($k \Rightarrow k+1$):}
Assume the claim holds for level $k$. We analyze four cases:

Case I: $\text{sign}(B, k) = \text{sign}(B^*, k)$ and  $\text{sign}(B, k+1) = \text{sign}(B^*, k+1)$.

Case II: $\text{sign}(B, k) = \text{sign}(B^*, k)$ and  $\text{sign}(B, k+1) \neq \text{sign}(B^*, k+1)$.

Case III: $\text{sign}(B, k) \neq \text{sign}(B^*, k)$ and  $\text{sign}(B, k+1) = \text{sign}(B^*, k+1)$.

Case IV: $\text{sign}(B, k) \neq \text{sign}(B^*, k)$ and  $\text{sign}(B, k+1) \neq \text{sign}(B^*, k+1)$.

For Case I, $B$ must include at least one link to reach level $k+1$, just as $B^*$ does. Therefore, $\vert B^*_{k+1}\vert \leq \vert B_{k+1}\vert$.  In Case II, $B$ must make at least one additional turn compared to $B^*$. Therefore, $\vert B^*_{k+1} \vert \leq \vert B_{k+1}\vert +1$.
Case III: by hypothesis, $B$ has, at least, one more turn than $B^*$ before reaching level $k$. As $B^*$ makes only one turn between levels $k$ and $k+1$, follows that $\vert B^*_{k+1} \vert \leq \vert B_{k+1}\vert$. Finally, Case IV can be proven using similar arguments to those presented for the previous cases. Figure \ref{fig:induction} illustrates Case II.
\begin{figure}[htb]
    \centering
    \includegraphics[scale=1]{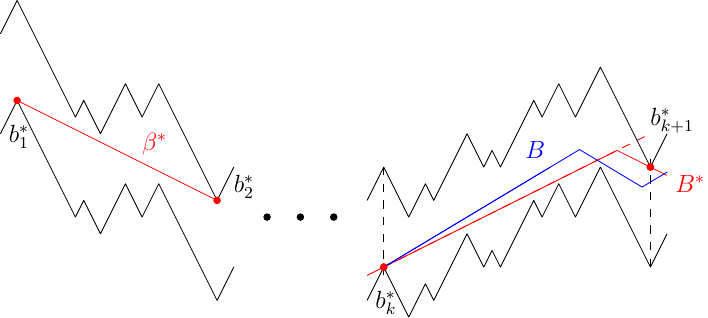}
    \caption{Proof of Theorem \ref{theo:min-links}.}
\label{fig:induction}
\end{figure}
\end{proof}

\begin{theorem}\label{theo:min-length}
     The $\beta^*$-path obtained using the GC algorithm minimizes the length among all polygonal paths that traverse the corridor $C(\alpha)$.
\end{theorem} 

\begin{proof}
Let $B$ be a \polygonal{\beta} within $C(\alpha)$ starting at a point $O$. Two consecutive links in $B$ alternate the sign of their slope and the length of $B$ can be computed using a straight lines starting at $O$ with the same slope than the first segment of $B$ as illustrated in Figure \ref{fig:path-length}.  
Thus, minimizing $\beta$ is equivalent to minimizing the length.
\begin{figure}[ht]
    \centering
    \includegraphics[scale=0.7]{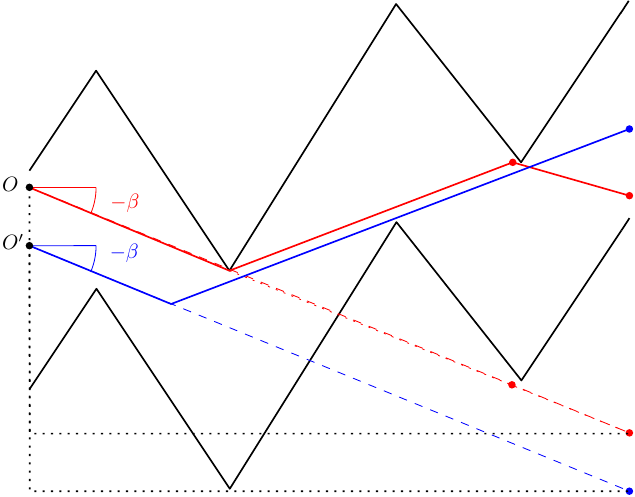}
    \caption{Two distinct \polygonal{\beta}s may have different numbers of links but the same total length, as the length depends solely on
    $\beta$.}
    \label{fig:path-length}
\end{figure}
\end{proof}

Theorems \ref{theo:min-links}, \ref{theo:min-length}, and \ref{theo:min-slope-complexity} support the main result of this work:

\begin{theorem} \label{theo:min-link-complexity}
The Min-Link and the Min-Length Problems  can be solved in $\Theta(n)$ time.
\end{theorem}
\begin{proof}
    Both Min-Link and Min-Length Problems are solved by computing the path given by the greedy construction in Section \ref{sec:min-link-path}. As shown in Theorem \ref{theo:min-slope-complexity}, $\beta^*$ can be computed in linear time; after that, the greedy construction will only make a constant number of operations on each level of the corridor (to check for collisions and select the next support reflex point). 
    
\end{proof}


\section{Conclusion and Future Research}\label{sec:conclusion}
In this work, we have presented an optimal algorithm to allow a ground vehicle to follow a drone at a bounded distance when both vehicles move along parallel lines within a vertical plane.
A geometric modeling of the problem enables the application of computational geometry tools to solve it efficiently. Three metrics were considered for minimization: the number of links, the slope, and the length of the ground vehicle's polygonal path.
The problem can be framed as a target tracking problem, which falls under a specialized class of surveillance problems. In our case, the trajectory and speed of the evader are known in advance, and the task is to compute the optimal path for the pursuer moving at a constant speed.

Extending this problem to more general scenarios, where both vehicles move in the plane through general curves, allows for a geometric formulation of a min-link problem within a 3D corridor. 
The problem statement is as follows: We are given a trayectory for the UAV determined by a polygonal curve in the plane and a constant speed. Compute a polygonal curve in the plane and a constant speed for the UGV so that the number of links is minimized while maintaining a distance at most $L$ to the UAV. A related optimization problem is to compute the minimum feasible speed for the UGV. The problem can be modelled as the computation of a polygonal curve traversing a ``$\alpha$-corridor" in 3D. The constant speeds of the vehicles force the orientation of the arms of both the corridor and the desired polygonal curve.

The min-link path problem is NP-hard for terrains \cite{kostitsyna2017complexity}. As a polyhedron in 3D is a terrain, the problem is hard for general 3D corridors. However, in our case, we deal with a special type of polyhedron, and the orientation of the links in the desired path is constant. Therefore, a possible line of research is to decide the complexity of the problem.  

An initial approach to tackling this 3D problem involves approximating it by decomposing the space into a bundle of planes and applying the 2D solution within each plane. As future work, we propose developing approximation algorithms to solve the problem more comprehensively.

\subparagraph*{Acknowledgments.} We thank A. Kasiuk, M.A. Pérez-Cutiño and J. Valverde for participating in early versions of this work, and all other participants of the II Soldrone Workshop, September 2024, Barbate, Spain, for helpful discussions.

\bibliographystyle{plain}
\bibliography{references}

\end{document}